\documentclass[a4paper,11pt]{article}
\usepackage[margin=2.5cm]{geometry}
\usepackage{graphicx}
\usepackage{epsfig}
\usepackage{amsmath}
\usepackage{amssymb}
\usepackage{cite}
\usepackage{float}
\usepackage{cancel}
\usepackage{amsfonts}
\usepackage{enumitem}
\usepackage[font={footnotesize,it}]{caption}
\usepackage{authblk}

\begin{document}

\begin{titlepage}

\title{Covariant Holographic Entanglement Negativity}

\author[1]{Pankaj Chaturvedi \thanks{\noindent E-mail:~  cpankaj1@gmail.com,\\ Present Address: Yau Mathematical Sciences Center, Tsinghua University, Beijing, China}}
\author[2]{Vinay Malvimat\thanks{\noindent E-mail:~ vinaymm@iitk.ac.in }}
\author[3]{Gautam Sengupta\thanks{\noindent E-mail:~  sengupta@iitk.ac.in}}

\affil[1,2,3]{
Department of Physics\\

Indian Institute of Technology\\ 

Kanpur, 208016\\ 

India}

\maketitle

\abstract{\noindent

We propose a covariant holographic conjecture for the entanglement negativity of mixed states in bipartite systems described by $d$-dimensional conformal field theories dual to bulk non static $AdS_{d+1}$ configurations. Application of our conjecture to $(1+1)$-dimensional conformal field theories dual to bulk rotating BTZ black holes exactly reproduces the corresponding entanglement negativity in the large central charge limit and characterizes the distillable entanglement. We further demonstrate that our conjecture applied to the case of bulk extremal rotating BTZ black holes also characterizes the entanglement negativity for the chiral half of the corresponding zero temperature  $(1+1)$-dimensional holographic conformal field theories. 

}

\end{titlepage}
\tableofcontents
\pagebreak
\section{Introduction}

Quantum entanglement has emerged as one of the central issues in the subject of quantum information theory in recent times. This has inspired significant attention towards the characterization of entanglement in extended quantum many body systems. In this context the entanglement entropy has been established as one of the crucial entanglement measures for bipartite quantum systems. To describe this it is necessary to consider the bipartition of a
quantum system into the subsystem-$A$ and its complement $A^c$ describing the the rest of the system. In such a scenario the entanglement entropy of the subsystem $A$ is given by von Neumann entropy of the reduced density matrix
as follows
\begin{equation}\label{s11}
 S_{A}= -Tr(\rho_{A}\log\rho_{A}),
\end{equation}
where $\rho_{A}$ is the reduced density matrix obtained by tracing out the degrees of freedom of $A^c$ i.e $\rho_{A}=Tr_{A^c}(\rho)$. The issue of characterizing the entanglement entropy for an extended quantum many body system is extremely non trivial as it involves the determination of the eigenvalues of an infinite dimensional density matrix-$\rho_{A}$. However
this issue is tractable in a $(1+1)$-dimensional conformal field theory 
$( CFT_{1+1} )$ through the {\it replica technique} as was demonstrated by Calabrese and Cardy in \cite{Calabrese:2004eu,Calabrese:2009qy}. This involves the description of the quantity $Tr(\rho^n_{A})$ in terms of the partition function  on a $n$-sheeted Riemann surface (${\cal Z}_n(A)$) with branch points at the boundaries between the subsystems $A$ and $A^c$ \cite{Calabrese:2004eu}. Then the entanglement entropy of the subsystem $A$ in the $CFT_{1+1}$ is given as follows 
\begin{equation}
S_{A}=-\frac{\partial}{\partial n} \log\left[Tr\left(\rho_A^n\right)\right]\Bigr|_{n=1}=\log{\cal Z}-\frac{\partial}{\partial n}\log{\cal Z}_n(A)\Bigr|_{n=1}. \label{EEodn}
\end{equation}
Here, $ {\cal Z} $ corresponds to the partition function of $CFT_{1+1}$ on a single sheet of the $n$-sheeted Riemann surface. The quantity ${\cal Z}_{n}(A)$ in the above equation is related to the two point function of certain branch-point twist fields in the corresponding $CFT_{1+1}$ which may then be computed in a straight forward fashion to obtain the entanglement entropy.  For time dependent states in a $CFT_{1+1}$ the reduced density matrix $\rho_A$ in eq.(\ref{EEodn}) has to be replaced by the time dependent  $\rho_{A}(t)=Tr_{A^c}\big({\rho}(t)\big)$ where the full density matrix $\rho(t)$ evolves according to the well known von-Neumann equation. Naturally this leads to a time dependent two point function of the twist and the anti/twist fields which describes the evolution of the entanglement entropy in the $CFT_{1+1}$\cite{Calabrese:2005in}. This remarkable progress in studying the time evolution of entanglement entropy in a $CFT_{1+1}$ has inspired focused attention in diverse areas such as quantum quenches, thermalization and quantum phase transitions \cite{ 1742-5468-2007-10-P10004, PhysRevA.78.010306,Calabrese:2009nk}. 

It is important to emphasize here that the entanglement entropy described above ceases to be a viable entanglement measure for mixed quantum states 
\footnote{Note that the system may be in a mixed state in a variety of physical situations. A  state may be mixed as a result of the interaction between the system and its environment or in the finite temperature case due to the interaction between the system and the heat reservoir. It may also occur if the system is in a degenerate ground state as in the case of a CFT which is dual to an extremal $AdS$ black hole. We will elaborate more about this case in one of the forthcoming sections.}. This relates to an important issue in quantum information theory termed as {\it purification} which involves the embedding of the bipartite system in a mixed state  within a larger system in a pure state. Through this procedure Vidal and Werner in a seminal work \cite{PhysRevA.65.032314} proposed a novel computable measure termed {\it entanglement negativity} which characterized the upper bound on the distillable entanglement for a mixed state. Consider a tripartite system  described by the subsystems  $A_1,A_2$ such that $A=A_1 \cup A_2$ and the rest of the system denoted by $A^c$, then the entanglement negativity of the subsystem $A_1$ may be defined as follows
\begin{equation}\label{s12}
{\cal E} \equiv \ln \big[Tr\mid\rho_{A}^{T_{2}}\mid\big].
\end{equation}
Where $\rho_{A_1 \cup A_2}=Tr_{A^c}(\rho)$ is the reduced density matrix of the subsystem $A=A_1 \cup A_2$ and the super-script $T_2$ corresponds to the operation of the partial transpose over the subsystem $A_2$. In this regard, if $|q_i^1\big>$ and $|q_i^2\big>$ represent the bases of Hilbert space corresponding to the subsystems $A_1$ and $A_2$ respectively, then the partial transpose with respect to $A_2$ degrees of freedom is expressed as 
\begin{equation}\label{trace}
\big<q_i^1q_j^2|\rho_{A_1 \cup A_2}^{T_{2}}|q_k^1q_l^2\big> = \big<q_i^1q_l^2|\rho_{A_1 \cup A_2}|q_k^1q_j^2\big>,
\end{equation}
Furthermore,  the entanglement negativity was shown to obey certain important properties such as monotonicity and non convexity in \cite{Plenio:2005cwa} . Note that similar to the case of entanglement entropy, obtaining the entanglement negativity for extended quantum systems is extremely complex as this also involves the evaluation of the eigenvalues of an infinite dimensional density matrix. However as in the case of the entanglement entropy,
the entanglement negativity for $CFT_{1+1}$ may be obtained through a variant of the replica technique \cite{Calabrese:2012ew,Calabrese:2012nk,Calabrese:2014yza}. As earlier this involves the computation of the correlation functions of branch point twist fields. Remarkably this leads to the upper bound on the distillable entanglement for the $CFT_{1+1}$ as anticipated from quantum information theory.

In an interesting communication Ryu and Takayanagi proposed a conjecture for the entanglement entropy of a subsystem in $d$-dimensional holographic CFTs
in the context of the $AdS/CFT$ correspondence \cite{Ryu:2006ef, Ryu:2006bv}.
Their conjecture relates the entanglement entropy $S_A$ for a region $A$ (enclosed by the boundary $\partial A$) in $(d)$-dimensional holographic
CFTs to the area of co-dimension two bulk $AdS_{d+1}$ static minimal surfaces (denoted by $\gamma_A$) anchored on the subsystem as follows
\begin{equation}\label{EEI}
S_A=\frac{Area(\gamma_A)}{(4G^{(d+1)}_N)}.
\end{equation}
Here, $G_N^{(d+1)}$ is the gravitational constant of the bulk space time. This conjecture has led to significant insights in exploring various aspects of  quantum entanglement in higher dimensional CFTs described in \cite{Takayanagi:2012kg, Nishioka:2009un} and references therein. It is important to note however that the Ryu-Takayanagi conjecture is applicable only to holographic $CFT_d$s dual to bulk static $AdS_{d+1}$ configurations. The reason for this is intricately related to the ambiguity in defining static minimal surfaces in non static configurations with a Lorentzian signature on which we will elaborate in a later section. Naturally this leads to the critical issue of the characterization of the holographic entanglement entropy for $CFT_d$s dual to non static $AdS_{d+1}$ configurations. In this context the authors Hubeny et. al in \cite{Hubeny:2007xt} have advanced a covariant holographic conjecture for the entanglement entropy of holographic $CFT_d$s dual to such non static $AdS_{d+1}$ configurations. This HRT-conjecture involves the light-sheet construction for the covariant entropy bound due to Bousso 
\cite{Bousso:1999xy, Bousso:1999cb, Bousso:2002ju}. The light sheet construction provides an elegant method for the characterization of a co-dimension two spacelike surface whose area provides a natural bound on the entropy flux. The explicit realization of this covariant holographic conjecture for the entanglement entropy has led to intense investigations for time dependent scenarios in holographic CFTs and has led to significant insights into issues of  quantum quenches and thermalization \cite{Balasubramanian:2011ur, Hartman:2013qma, Albash:2010mv, Caputa:2013eka, Mandal:2016cdw, David:2016pzn}.  

The above discussion naturally led to the crucial issue of establishing a precise and elegant holographic prescription
for the entanglement negativity of mixed states in conformal field theories at finite temperatures in the $\it{AdS}/\it{CFT}$ framework. A recent conjecture for such a holographic prescription for the entanglement negativity was proposed by us (CMS) in \cite{Chaturvedi:2016rcn,Chaturvedi:2016rft}. It could be demonstrated that the holographic entanglement negativity (${\cal E}$) of a subsystem-$A$ is proportional to a particular algebraic sum of the co-dimension two static minimal surfaces in the dual bulk $(d+1)$-dimensional $AdS$ space time  which reduces to the sum of holographic mutual information between different subsystems as follows
\begin{eqnarray}
{\cal E}&=&\lim_{B\rightarrow A^c}\frac{3}{4}\big[{\cal I}(A,B_1)+{\cal I}(A,B_2)\big] \label{s15}\\ 
{\cal I}(A,B_i)&=& S_{A}+S_{B_i}-S_{A\cup B_i}=\frac{1}{4G_{N}^{(d+1)}}({\cal A}_{A}+{ \cal A}_{B_i}-{\cal A}_{A\cup B_i})\ i=\{1,2\}\nonumber, 
\end{eqnarray}
 Where ${\cal A_{\gamma}}$ is the area of the minimal surface anchored on the corresponding subsystem $\gamma=\{A,B_1,B_2,A\cup B_1, A\cup B_2\}$ and the limit $B\rightarrow A^c$ in the above expression corresponds to extending the two large finite subsystems $B_1$ and $B_2$ on either side of ($A$) to infinity such that in this limit $B_1\cup B_2$ denoted as $B$ is the rest of the system $A^c$.    

In the $AdS_3/CFT_2$ scenario, we employed our conjecture to compute the entanglement negativity of an extended bipartite system described by a finite temperature $CFT_{1+1}$  from the algebraic sum of geodesics in the bulk which is an Euclidean BTZ black hole. We demonstrated that the holographic entanglement negativity obtained from our conjecture matches exactly with the $CFT_{1+1}$ result in the large central charge limit. Furthermore, in an another communication \cite{Chaturvedi:2016rft} we applied our conjecture to obtain holographic entanglement negativity of $d$-dimensional CFTs at finite temperatures which are dual to bulk $\it{AdS_{d+1}}$-Schwarzschild black holes. Remarkably it could be demonstrated that in both the cases the holographic entanglement negativity precisely captured the distillable quantum entanglement at all temperatures. This naturally constituted a strong evidence for the universality of our conjecture.

Notice however that our holographic conjecture proposed and studied in \cite {Chaturvedi:2016rcn,Chaturvedi:2016rft}
is only applicable to conformal field theories which are dual to static bulk $AdS$ black holes due to the ambiguity in the definition of minimal surfaces in non static space time geometries. In this article we propose such a holographic conjecture for the covariant entanglement negativity of mixed states for bipartite systems described by $d$-dimensional conformal field theories dual to bulk non static $AdS_{d+1}$ gravitational configurations. Subsequently using our conjecture, we obtain the covariant entanglement negativity of a single interval in a $CFT_{1+1}$ at a finite temperature dual to a bulk stationary $AdS_3$ configuration involving a  rotating BTZ black hole in the $AdS_3/CFT_2$ scenario. Furthermore we also apply our conjecture to obtain the same for a $CFT_{1+1}$ dual to an
extremal rotating bulk BTZ black hole. In both of these case we also compute the entanglement negativity of the corresponding $CFT_{1+1}$ by employing the replica technique and demonstrate that it matches exactly with the bulk result obtain using our conjecture in the large central charge $c$ limit. Furthermore as earlier for static bulk configurations, it is observed that the holographic entanglement negativity precisely leads to the distillable quantum entanglement.

This article is organized as follows. In section-2 we review the HRT prescription for the covariant holographic entanglement entropy in the $AdS_{d+1}/CFT_d$ scenario. In section-3 we propose our covariant holographic conjecture for the entanglement negativity of  $d$ dimensional CFTs which are dual to non static $AdS_{d+1}$ space times.  In section-4 we use our prescription to compute the entanglement negativity of a $CFT_{1+1}$ dual to a rotating non-extremal BTZ black hole background . In section-5 we compute the entanglement negativity of a finite temperature $CFT_{1+1}$ with angular momentum and demonstrate that the large-c limit of this result matches exactly with that obtained from bulk using our holographic prescription. In section-6 we compute the covariant holographic entanglement negativity of the $CFT_{1+1}$ dual to the extremal rotating BTZ black hole and show that entanglement negativity captures the distillable quantum entanglement at all temperatures. Subsequently, in section-7 we compute the entanglement negativity in the dual CFT and show that the result once again matches exactly with the bulk computation in the large central charge ($c$) limit. In section-8 we summarize our results and conclude.

\section{Review of the covariant holographic entanglement entropy}

As mentioned earlier the replica procedure provided by Calabrese et al. is a systematic method to obtain the entanglement entropy of a subsystem in a $CFT_{1+1}$. As there is no such direct procedure to evaluate the entanglement entropy in the higher dimensional CFTs one must make use of the holographic conjecture by Ryu and Takayanagi which we briefly described in the introduction. Using their holographic prescription , it was possible to obtain the entanglement entropy of various higher dimensional holographic CFTs (see\cite{Cadoni:2009tk,Tonni:2010pv,Fischler:2012ca,Fischler:2012uv,Blanco:2013joa,PhysRevD.94.066004,Kundu:2016dyk} and references therein). However unlike the replica procedure which is applicable to both time dependent and time independent scenarios in $CFT_{1+1}$, the Ryu and Takayanagi conjecture is valid only for the CFTs which are dual to static $AdS$ space time backgrounds.  The reason for this is as follows. The Ryu and Takayanagi conjecture relates the entanglement entropy of a subsystem in $CFT_d$ to a minimal surface in the corresponding bulk $AdS_{d+1}$ space time.  These static minimal surfaces are well defined in time independent scenarios as it is always possible to perform a Wick rotation leading to an Euclidean $AdS$ geometry in the bulk. However for describing time dependent states in a boundary CFT, the corresponding dual bulk involves a non-static $AdS$ spacetime with a Lorentzian geometry in which case one may always contract the spacelike surface along the time direction to reduce the area of the minimal surface to arbitrarily small values.   Therefore, the Ryu-Takayanagi conjecture is unsuitable for describing the time evolution of the entanglement entropy in CFTs. This issue was resolved by Hubeny, Rangamani and Takayanagi(HRT) in\cite{Hubeny:2007xt} where the authors provided a covariant prescription for the holographic entanglement entropy of a $d$-dimensional CFT. We briefly review their proposal here.

The HRT conjecture for the covariant holographic entanglement entropy of a subsystem in a $CFT_d$ dual to a non-static $AdS_{d+1}$ space time\cite{Hubeny:2007xt}, was inspired by the light sheet construction for the covariant Bousso bound \cite{Bousso:1999xy,Flanagan:1999jp,Bousso:2002ju}. Consider a co-dimension two spacelike surface ${\cal S}$, a light sheet $L_{\cal S}$ for this ${\cal S}$ is defined by null geodesic congruences whose expansion is non-positive definite. According to the Bousso bound, the thermodynamic entropy $S_{L_{\cal S}}$ through a light sheet $L_{\cal S}$ is bounded by the area of ${\cal S}$ in Planck units
\begin{equation}\label{r21}
 S_{L_{\cal S}}\leq \frac{Area({\cal S})}{4G_N}.
\end{equation}

If $L_{-}$ and $L_{+}$ represent the past and the future light sheets respectively, then the corresponding null expansions denoted by $\theta_{\pm}$ obey the inequality $\theta_{\pm}\leq 0$. The authors in \cite{Hubeny:2007xt} argue that the holographic entanglement entropy saturates the above mentioned Bousso bound. They consider a $d$-dimensional strongly coupled conformal field theory living on the boundary of $(d+1)$-dimensional AdS space time. The asymptotic boundary of the $AdS_{d+1}$ space time where the conformal field theory is situated is divided into two time dependent regions $A_t$ and $A_t^c$. The boundary of the region $A_t$ is denoted as $\partial A_t$. Following this, the past and future light sheets of a for spacelike surface $\partial A_t$, denoted as $\partial L_{+}$ and $\partial L_{-}$ are constructed. One is then required to extend  $\partial L_{+}$ and $\partial L_{-}$ into the bulk in such a way that their extensions denoted by $ L_{+}$ and $L_{-}$ respectively, are also light sheets corresponding to a $(d-1)$-dimensional spacelike surface ${\cal Y} _t=L_+\cap L_-$ which is anchored to  $\partial A_t$. According to the HRT proposal, out of all such surfaces ${\cal Y}_t$ the holographic entanglement entropy of the region $A_t$ is given by the surface which has the minimum area (${\cal Y}_t^{min}$). The authors also showed that this surface (${\cal Y}_t^{min}$) is also the extremal surface (${\cal Y}^{ext}$) anchored to $\partial A_t$ and the null expansions for this spacelike surface vanish ( i.e $\theta_{\pm}=0$). 
\begin{equation}\label{r22}
 S_{A_t}=\frac{Area({\cal Y}_t^{min})}{4G_N^{(d+1)}}=\frac{Area({\cal Y}^{ext})}{4G_N^{(d+1)}}.
\end{equation}

The authors(HRT) verified their prescription in the $AdS_3/CFT_2$ scenario as follows. They employed the above mentioned holographic prescription to obtain the entanglement entropy of the CFT dual to the rotating BTZ black hole background. They then computed the entanglement entropy of a subsystem-$A$ in the corresponding dual $CFT_{1+1}$ through the standard replica technique and demonstrated that the result matches exactly with that obtained using their conjecture which is as follows
  \begin{equation}\label{r221}
  S_A=\frac{c}{6}\log\big[\frac{\beta_{+}\beta_{-}}{\pi^{2}a^2}\sinh{(\frac{\pi\ell}{\beta_+})}\sinh{(\frac{\pi\ell}{\beta_-}})\big].
 \end{equation}
 
 Where, $\beta_+=\beta(1+\Omega)$ and $\beta_-=\beta(1-\Omega)$ are the left and the right moving temperatures the $CFT_{1+1}$, $\Omega$ is the angular velocity, $\ell$ is the length of the subsystem-$A$ and $a$ is the UV cut-off for the field theory.

\section{Covariant holographic entanglement negativity conjecture }

In this section we propose a covariant prescription for the holographic entanglement negativity in the $AdS_{d+1}/CFT_{d}$ scenario. In order to do this we first briefly review our recently conjectured prescription for the holographic entanglement negativity of CFTs dual to static $AdS$ space time configurations \cite{Chaturvedi:2016rft}. The $d$-dimensional conformal field theory on the asymptotic boundary of a static $AdS_{d+1}$ space time is partitioned into two regions $A$ and it's complement $A^c$. The region $B$ within the complement $A^c$ consists of two large but finite disjoint subsystems $B_1$ and $B_2$ on either side of $A$, such that $B=B_1\cup B_2$. The holographic entanglement negativity which quantifies the entanglement between a subsystem-$A$ and the rest of the system $A^c$, is given by an algebraic sum of the areas of the static minimal surfaces in the bulk as follows
\begin{equation}\label{r23}
  {\cal E} = \lim_{B\rightarrow A^c}\frac{3}{16G_{N}^{(d+1)}} \big[2{\cal A}_{A}+{\cal A}_{B_1}+{\cal A}_{B_2}-{\cal A}_{A\cup B_1}-{\cal A}_{A\cup B_2}\big]. 
\end{equation}
Where ${\cal A}_\gamma$ represents the co dimension two static minimal surface anchored on the corresponding subsystem $\gamma$ and the limit $B\to A^c$ has to be interpreted as extending the subsystems $B_1$ and $B_2$ to infinity such that in this limit $B$ is $A^c$. Note that the above equation may be re-expressed in terms of the holographic mutual information denoted by ${\cal I}(A,B_1)$ and ${\cal I}(A,B_2)$ as given by eq.(\ref{s15}). 

However the above expression for the holographic entanglement negativity is valid only for $CFT_d$ dual to static $AdS_{d+1}$ gravitational configurations such as the $AdS_{d+1}$- Schwarzschild black holes. For a $CFT_d$  dual to a non static bulk $AdS_{d+1}$ gravitational configuration it is possible to reduce the area of the minimal surface to zero which leads to a degenerate situation. As reviewed in the previous section a similar issue also occurs for the case of the entanglement entropy and is resolved through the replacement of the area of the minimal surface with that of the extremal surface \cite {Hubeny:2007xt}. It is now clear that to obtain the covariant holographic entanglement negativity for a $CFT_d$ dual to a non static $AdS_{d+1}$ gravitational configuration, it is required to consider the areas of the extremal surfaces (${\cal Y}^{ext}$) instead of the areas of the static minimal surfaces (${\cal A}$) in  eq.(\ref{r23}). The above discussion therefore leads to the following expression for the covariant holographic entanglement negativity 
\begin{equation}\label{r24}
  {\cal E} = \lim_{B\rightarrow A^c}\frac{3}{16G_{N}^{(d+1)}} \big[2{\cal Y}_{A}^{ext}+{\cal Y}_{B_1}^{ext}+{\cal Y}_{B_2}^{ext}-{\cal Y}^{ext}_{A\cup B_1}-{\cal Y}^{ext}_{A\cup B_2}\big]. 
\end{equation}
where ${\cal Y}_{\gamma}^{ext}$  is the area of the extremal surface anchored on the corresponding subsystem $\gamma$. The above equation may once again be re-expressed as the holographic mutual information between pairs of subsystems $(A,B_1)$ and $(A,B_2)$ as
\begin{eqnarray}
{\cal E}&=&\lim_{B \to A^c}\left[\frac{3}{4}\big({\cal I}(A,B_1)+{\cal I}(A,B_2)\big)\right],\\
{\cal I}(A,B_1)&=& S_{A}+S_{B_1}-S_{A\cup B_1}=\frac{1}{4G_{N}^{(d+1)}}({\cal Y}_{A}^{ext}+{ \cal Y}_{B_1}^{ext}-{\cal Y}_{A\cup B_1}^{ext}),\\
{\cal I}(A,B_2)&=& S_{A}+S_{B_2}-S_{A\cup B_2}=\frac{1}{4G_{N}^{(d+1)}}({\cal Y}_{A}^{ext}+{ \cal Y}_{B_2}^{ext}-{\cal Y}_{A\cup B_2}^{ext}),
\end{eqnarray}

In the forthcoming sections we put our holographic conjecture to test in the $AdS_3/CFT_2$ scenarios. In the next section, we employ the above described holographic proposal of ours to evaluate the covariant holographic entanglement negativity of a mixed state in a bipartite system described by a $CFT_{1+1}$ dual to the bulk rotating non-extremal BTZ black hole background which is a stationary space time. In the subsequent section, we compute the same for the finite temperature $(1+1)$-dimensional CFT with rotation and demonstrate that it matches exactly with the bulk result in the large central charge limit.

\section{Covariant holographic entanglement negativity of a $CFT_{1+1}$ dual to a rotating non-extremal BTZ }
In this section we compute the covariant holographic entanglement negativity of a mixed state for an extended bipartite quantum system described by a $CFT_{1+1}$ dual to a rotating non-extremal BTZ black hole in the $AdS_3/CFT_2$ scenario(we will consider the extremal black hole case in a latter section as it is special). Note that this corresponds to a finite temperature $CFT_{1+1}$  with a conserved angular momentum. The metric for a rotating BTZ black hole is given by
\begin{equation}\label{s31}
 ds^{2}=-\frac{(r^2-r_{+}^2)(r^2-r_{-}^2)}{r^2}dt^2 + \frac{r^2 dr^2}{(r^2-r_{+}^2)(r^2-r_{-}^2)}+ r^2(d\phi-\frac{r_{+}r_{-}}{r^2}dt)^2.
\end{equation}

Here, we have set the the AdS length scale to unity ($R=1$) and $r_-$ and $r_+$ in the above equation correspond to the inner and outer horizon radius of the black hole.  The mass $M$, the angular momentum $J$, the Hawking temperature $T_H$ and the angular velocity $\Omega$ of the black hole may be expressed in terms of $r_-$ and $r_+$ as follows 
\begin{eqnarray}\label{s32}
 8G^{(3)}M = r_{+}^2+r_{-}^2, \hspace{0.5cm} J=\frac{r_{+}r_{-}}{4G^{(3)} }, \hspace{0.5cm} \beta=\frac{1}{T_H}=\frac{2\pi r_{+}}{r_{+}^2-r_{-}^2}, \hspace{0.5cm} \Omega=\frac{r_{-}}{r_{+}}, \hspace{0.5cm}\beta_{\pm}=\beta(1\pm\Omega).
\end{eqnarray}

Notice that for a BTZ black hole $\phi$ is periodic i.e $\phi\sim\phi+2\pi$ where as for a BTZ black string  $\phi\in \mathbb{R}$. For the case of a spatially non compact $CFT_{1+1}$  which we are interested in, it is required to consider the bulk dual as BTZ black string. Note that the metric in eq.(\ref{s31}) is a stationary gravitational configuration and hence as discussed in the section-3 the covariant proposal of HRT must be employed to compute the holographic entanglement entropy. We  briefly review their covariant proposal \cite{Hubeny:2007xt} which involves the following co-ordinate transformation as a first step,
\begin{eqnarray}
 w_{\pm}&=&\sqrt{\frac{r^2-r_{+}^2}{r^2-r_{-}^2}} e^{\frac{2\pi}{\beta_{\pm}}(\phi\pm t)}\equiv X\pm T,\label{s33}\\
 Z&=&\sqrt{\frac{r_{+}^2-r_{-}^2}{r^2-r_{-}^2}} e^{(r_{+}\phi_{i}~-~tr_{-})}\label{s33i}.
\end{eqnarray}

Where $w_{\pm}= X\pm T$ are the light cone coordinates and  $(X, T,Z)$ are the the Poincar\'{e} coordinates. The above transformation maps the metric of the BTZ black hole given in eq.(\ref{s31}) to the Poincar\'{e} metric of the pure $AdS_{3}$ given by
\begin{equation}\label{s34}
 ds^2=\frac{dw_{+}dw_{-}+dZ^2}{Z^2}\equiv\frac{-dT^2+dX^2+dZ^2}{Z^2}.
\end{equation}

The length of the required spacelike geodesics anchored on the boundary subsystems is well known for the pure $AdS_3$ space time and leads to the following expression \footnote{Note that for the present case of the rotating BTZ black hole which is stationary spacetime the the required extremal surface turns out to be a spacelike geodesic. However for a general non-static non-stationary bulk configuration the entanglement entropy is given by the extremal surface which is to be determined by setting the null expansions to zero($\theta_{\pm}=0$) as described earlier.}
\begin{eqnarray}
 {\cal L}_{\gamma}&=&\log[\frac{(\Delta \phi)^2}{\varepsilon_{i}\varepsilon_{j}}],\label{s351}\label{s351}\\
 \varepsilon_{i} &=& \sqrt{\frac{r_{+}^2-r_{-}^2}{r_{\infty}^2}}e^{(r_{+}\phi_{i}~-~t_0 r_{-})}\label{s352}.
\end{eqnarray}
Where  ${\cal L}_{\gamma}$ represents the length of spacelike geodesic anchored to the boundary of the subsystem-$\gamma$ which is a spacelike interval $[\phi_{i},\phi_{j}]$ and $r_{\infty}$ is the infrared cut-off for the bulk BTZ black hole background where as $\varepsilon_{i}$ is the same for the pure $AdS_3$ space time. Notice that the constant time slice is taken along $t=t_0$. Upon re-expressing eq.(\ref{s351}) in terms of BTZ coordinates then using the Ryu-Takaynagi formula and the Brown-Hennaux formula ($c=\frac{3 R}{2 G^{(3)}}$) the following expression for the entanglement entropy was obtained \cite{Hubeny:2007xt} 
\begin{equation}\label{s36}
  S_{\gamma} = \frac{c}{6}\log\big[\frac{\beta_{+}\beta_{-}}{\pi^2a^2}\sinh{\big(\frac{\pi\mid\phi_{i}-\phi_{j}\mid}{\beta_{+}}\big)}\sinh{\big(\frac{\pi\mid\phi_{i}-\phi_{j}\mid}{\beta_{-}}\big)}\big],
\end{equation}
where $S_{\gamma}$ represents the entanglement entropy of a subsystem-$\gamma$, $a$ is the UV cut-off for the boundary CFT related to the bulk infrared cut-off($a\sim\frac{1}{r_{\infty}}$).

Having briefly reviewed the method to obtain the expression for the holographic entanglement entropy, we now employ our holographic conjecture to determine the covariant holographic entanglement negativity for a $CFT_{1+1}$ dual to a rotating non-extremal BTZ black hole. As discussed earlier the extremal surfaces in eq.(\ref{r24}) for the covariant holographic entanglement negativity are the space-like geodesics for the present case and hence, the entanglement negativity is given as follows
\begin{eqnarray}\label{s37}
{\cal E} = \lim_{B\rightarrow A^c}\frac{3}{16G_{N}^{(d+1)}} \big[2{\cal L}_{A}+{\cal L}_{B_1}+{\cal L}_{B_2}-{\cal L}_{A\cup B_1}-{\cal L}_{A\cup B_2}\big].
\end{eqnarray}
In order to compute the entanglement negativity of the subsystem $A$ we make the following identification for the points $(\phi_1,\phi_2,\phi_3,\phi_4)\equiv(-L,u,v,L)$ which implies that the required subsystems $A$, $B_1$, $B_2$ correspond to the intervals given by $[u,v]$, $[-L,u]$ and $[v,L]$ respectively. We use the expression given by eq.(\ref{s351}) to obtain the required lengths of the geodesics which are anchored to each of these intervals i.e ${\cal L}_A$, ${\cal L}_{B_i}$ and ${\cal L}_{A\cup B_i}$ $(i=1,2)$. These are then substituted in eq.(\ref{s37}) and the limit $B \to A^c$ (which corresponds to the limit $L\to\infty$) is taken to obtain the entanglement negativity  as
\begin{equation}\label{s38}
{\cal E}= \frac{c}{4}\log\big[\frac{\beta_{+}\beta_{-}}{\pi^2a^2}\sinh{\big(\frac{\pi\ell}{\beta_{+}}\big)}\sinh{\big(\frac{\pi\ell}{\beta_{-}}\big)}\big]-\frac{\pi c\ell}{2\beta(1-\Omega^2)}.
\end{equation}

Re-expressing the above equation in terms of the entanglement entropy  and the thermodynamic entropy of subsystem-A, we get
\begin{equation}\label{s39}
 {\cal E}= \frac{3}{2}\bigg[S_A-S^{th}_A\bigg].
\end{equation}

Remarkably the above result clearly demonstrates that the covariant holographic entanglement negativity leads to the distillable quantum entanglement by eliminating the contribution from the thermal correlations. In the next section we compute the entanglement negativity of a mixed state for a bipartite system described by the corresponding dual $CFT_{1+1}$ using the replica procedure and demonstrate that it matches exactly with the above result in the large central charge limit.

\section{Entanglement negativity in a $CFT_{1+1}$ with angular momentum and at a finite temperature }
In this section, we compute the entanglement negativity of a bipartite quantum system in a $(1+1)$-dimensional CFT with a conserved angular momentum and at a finite non-zero temperature. We will demonstrate that the entanglement negativity precisely captures the distillable entanglement and also show that in the large-$c$ limit it matches exactly with the result obtained in the previous section using our covariant holographic conjecture. As described in the introduction a variant of the replica technique was developed by Calabrese, Cardy and  Tonni in \cite{Calabrese:2014yza} to evaluate the entanglement negativity of a mixed state for a bipartite quantum system described by a $(1+1)$-dimensional finite temperature conformal field theory, which is expressed as
\begin{equation}
{\cal E}=\lim_{n_e \to 1}\log(Tr[(\rho_A^{T})^{n_e}]).
\end{equation}
The full system is partitioned into subsystem-$A$ which is an interval $[u,v]$ and the rest of the system denoted as $A^c$. The quantity $\rho_A^{T}$ is the reduced density matrix of the subsystem-$A$ partial transposed over $A^c$ ( For the details of how to obtain the above expression for the entanglement negativity of a bipartite quantum system from its definition for the tripartite configuration given by eq.(\ref{s12}),  see \cite{Calabrese:2014yza} or the appendix of \cite{Chaturvedi:2016rft}).  Notice that this definition is valid only when the parity of $n$ is even ($n=n_e$) i.e the above definition has to be interpreted as an analytic continuation of an even sequence in $n_e$ to $n_e\to1$. The authors in \cite{Calabrese:2014yza} demonstrated that for a $CFT_{1+1}$ at a finite temperature the quantity $(Tr[(\rho_A^{T})^{n_e}])$ is related to a particular four point function of the twist and the anti-twist fields, resulting in the following expression for the entanglement negativity

\begin{equation}\label{r19}
  {\cal E}=\lim_{L \to \infty}\lim_{n_e \to 1}\log\left[\big<{\cal T}_{n_e}(-L)\overline{{\cal T}}^2_{n_e}(u){\cal T}^2_{n_e}(v)\overline{{\cal T}}_{n_e}(L)\big>_{\beta}\right],
\end{equation}

where $<...>_\beta$ in the above equation indicates that this four point function has to be evaluated in the finite temperature $CFT_{1+1}$ on an infinite cylinder of circumference $\beta$ ( $\beta=1/T$ where $T$ is the temperature ). In the above equation ${\cal T}_{n_e}$ and $\overline{{\cal T}}_{n_e}$ are the twist and the anti-twist fields which are primary operators with the scaling dimension $\Delta_{n_e}$, whereas ${\cal T}^2_{n_e}$ and $\overline{{\cal T}}^2_{n_e}$ are the twist fields which are primary operators that connect $j$-th sheet of the Riemann surface with $(j+2)$-th sheet and their scaling dimension $\Delta_{n_e}^{(2)}$ are given as follows
\begin{eqnarray}
\Delta_{n_e}&=&\frac{c}{12}(n_e-\frac{1}{n_e}),\\
\Delta_{n_e}^{(2)} &=&2\Delta_{\frac{n_e}{2}}= \frac{c}{6}\left(\frac{n_e}{2}-\frac{2}{n_e}\right),
\end{eqnarray}
The authors in \cite{Calabrese:2014yza} computed the entanglement negativity using the above eq.(\ref{r19}) to illustrate that this quantity is a precise measures of the distillable quantum entanglement through the removal of the thermal contributions.

 In present case we consider a $CFT_{1+1}$ at a finite temperature, having a conserved angular momentum. The partition function of this CFT is given by
\begin{equation}\label{s21}
 Z(\beta)=Tr(e^{-\beta (H-i\Omega_{E}J )})=Tr(e^{-\beta_{+}L_0-\beta_{-}\bar{L}_0}).
\end{equation}

Where, $\beta$ is the inverse temperature, $J$ is the angular momentum, $\Omega_{E}$ is the Euclidean angular velocity (related to the Minkowskian angular velocity by  $\Omega_{E}=-i\Omega$). In the above equation we have identified the left and right moving temperatures as $\beta_{\pm}=\beta(1\pm i\Omega_{E})$. Notice that the conserved Virasoro charges are the Hamiltonian $H=L_0+\bar{L}_0$ and the angular momentum $J=L_0-\bar{L}_0$ with $L_0$ and $\bar{L}_0$ being the holomorphic and anti-holomorphic zeroth mode Virasoro generators. Note that this $CFT_{1+1}$ lives on a twisted cylinder and may be obtained by the Euclidean CFT on a complex plane by the following conformal transformation 
\begin{equation}\label{s22}
 w= \frac{\beta(1-i\Omega_{E})}{2\pi}\log[z],
\end{equation}
where $z$ denotes the coordinates on the complex plane and $w$ corresponds to the coordinates on the twisted cylinder mentioned above. This leads to the following expression for the entanglement negativity of an extended bipartite quantum system ($A\cup A^c$) in a $CFT_{1+1}$ at finite temperature and with a conserved angular momentum.
\begin{equation}\label{s23}
{\cal E}= \lim_{L\rightarrow\infty}\lim_{n_{e}\rightarrow1}\log[\big<{\cal T}_{n_{e}}(-L)\bar{\cal T}^{2}_{n_{e}}(u){\cal T}^{2}_{n_{e}}(v)\bar{\cal T}_{n_{e}}(L)\big >_{\beta,\Omega_{E}}].
\end{equation}

In the above expression the subscript $\beta,\Omega_{E} $ indicates that the four point function has to be evaluated in a $CFT_{1+1}$ on a twisted cylinder. The four point function on a twisted cylinder may be obtained from the four point function on the complex plane, by using the following transformation
\begin{equation}\label{s26}
 \big<{\cal T}_{n_e}(w_1)\overline{{\cal T}}^2_{n_e}(w_2){\cal T}^2_{n_e}(w_3)\overline{{\cal T}}_{n_e}(w_4)\big>_{\beta,\Omega_{E}}=\prod_{i=1}^{4}\mid\frac{dz_i}{dw_i}\mid^{\Delta^{i}_{n_{e}}} \big<{\cal T}_{n_e}(z_1)\overline{{\cal T}}^2_{n_e}(z_2){\cal T}^2_{n_e}(z_3)\overline{{\cal T}}_{n_e}(z_4)\big>_{\mathbb{C}}.
\end{equation}

 $\Delta^{i}_{n_{e}}$ in the above equation represents the scaling dimension of the operator at $z_i$.
The four point function in a $CFT_{1+1}$ on the complex plane may be shown to have the following form  \cite{Calabrese:2014yza}
\begin{equation}\label{s25}
\big<{\cal T}_{n_e}(z_1)\overline{{\cal T}}^2_{n_e}(z_2){\cal T}^2_{n_e}(z_3)\overline{{\cal T}}_{n_e}(z_4)\big>_{\mathbb{C}}=\frac{c_{n_e}c^{(2)}_{n_e}}{z_{14}^{2\Delta_{n_e}}z_{23}^{2\Delta^{(2)}_{n_e}}}\frac{{\cal F}_{n_e}(x)}{x^{\Delta^{(2)}_{n_e}}},~~~~~x\equiv\frac{z_{12}z_{34}}{z_{13}z_{24}}.
\end{equation}

Notice from above that the four point function is determined only up to a function of the cross-ratio ($x=\frac{z_{12}z_{34}}{z_{13}z_{24}}$) and this function denoted here as ${\cal F}_{n_e}(x)$ can not be fixed by the conformal symmetry alone as it depends on the full operator content of the theory. We compute the required four point function on the twisted cylinder by substituting the four point function on a plane given by eq.(\ref{s25}) in the  transformation given by eq.(\ref{s26}). We also identify the points $(w_1,w_2,w_3,w_4)\equiv(-L,u,v,L)$. After obtaining the four point function function, it may then be substituted in eq.(\ref{s23}) to obtain the entanglement negativity as follows
\begin{equation}\label{s27}
{\cal E}= \frac{c}{4}\log\big[\frac{\beta_{+}\beta_{-}}{\pi^{2}a^2}\sinh{(\frac{\pi\ell}{\beta_+})}\sinh{(\frac{\pi\ell}{\beta_-}})\big]-\frac{\pi c\ell}{2\beta(1+\Omega_{E}^2)}+f(e^{-\frac{2\pi \ell}{\beta(1+\Omega_{E}^2}})+\ln[c^2_{1/2} c_1],
 \end{equation}
where $\ell=\mid u-v\mid$ is length of the subsystem-$A$ and $a$ is UV cut-off for the field theory, $c_1$ and $c_{\frac{1}{2}}$ are normalization constants of the two point function of the twist and the anti-twist fields appropriately. The function $f(x)$ in the above equation is defined in the replica limit as follows
\begin{equation}
f(x)=\lim_{n_e \to 1}\ln[{\cal F}_{n_e}(x)],~~~~~\lim_{L \to \infty}x=e^{-\frac{2\pi \ell}{\beta(1+\Omega_{E}^2)}}.
\end{equation}

It is to be noted that this function $f(x)$ is undetermined except at the limiting cases $x=0$ and $x=1$.  The value for this function at these two limits is given by
\begin{equation}\label{r111}
 f(1)=0, \hspace{0.5cm} f(0)=\lim_{n_e\rightarrow1}\ln[\frac{C_{{\cal T}_{n_e}\bar{{\cal T}}_{n_e}^2\bar{{\cal T}}_{n_e}}}{c_{n}^{(2)}}],
\end{equation}
These expressions for the limiting cases of the function $f(x)$ follow from the argument provided in  \cite{Calabrese:2014yza} for the finite temperature scenario. The constants $C_{{\cal T}_{n_e}\bar{{\cal T}}_{n_e}^2\bar{{\cal T}}_{n_e}}$  and $c_{n}^{(2)}$ in the equation above are the  coefficients of the leading term in the operator product expansion (OPE) of  ${\cal T}_{n_e}(z_1)\bar{{\cal T}}_{n_e}^2(z_2)$ and ${\cal T}_{n_e}^2(z_1)\bar{{\cal T}}_{n_e}^2(z_2)$ respectively. Following this we make a Wick rotation $\Omega=i \Omega_{E}$  and  re-write the right hand side of eq.(\ref{s27}) in terms of entanglement entropy($S_A$) given in eq.(\ref{r221}) and the thermodynamic entropy of subsystem $A$ ($S_A^{th}= \frac{\pi c\ell}{3\beta(1-\Omega^2)}$) which leads us to 
\begin{equation}\label{s28}
 {\cal E}=\frac{3}{2}[S_{A}-S^{th}_{A}]+f(e^{-\frac{2\pi \ell}{\beta(1-\Omega^2)}})++\ln[c^2_{1/2} c_1].
\end{equation}

We see from the expression above that the  entanglement negativity  serves as a precise measure of the quantum distillable entanglement at finite temperatures through the elimination of the classical/thermal contributions. 
Notice that for the angular velocity $\Omega=0$, the above equation reduces to the expected result for the entanglement negativity of a finite temperature $CFT_{1+1}$ without angular momentum as obtained in \cite{Calabrese:2014yza}.
In the large central charge limit of the $CFT_{1+1}$ as argued in our earlier work \cite{Chaturvedi:2016rft} the first two universal terms
are $O[c]$ and dominate over the non-universal term involving the function $f(x)$ and the constant in eq.(\ref{s28}) are expected to be subleading and of order $O (\frac{1}{c})$. The detailed argument for this issue involves results from the large-$c$ limit of the conformal block expansions for the correlations functions computed through monodromy techniques \cite {Hartman:2013mia,Fitzpatrick:2014vua, Kulaxizi:2014nma} \footnote {We have recently completed the relevant computations for the four point function of the twist fields describing the  entanglement negativity, through these techniques which provide extremely strong evidence in favor of this assertion. This result is in preparation for communication.}. The above arguments clearly indicate that
the subleading terms in eq.(\ref{s28}) involving the undetermined function $f(x)$ may be neglected in the large central charge limit. This leads us to the following expression for the large-$c$ limit of the entanglement negativity of the mixed state descried by the single interval in a finite temperature $CFT_{1+1}$ with a conserved charge described by the bulk angular momentum.
\begin{equation}\label{s29}
 {\cal E}= \frac{3}{2}[S_{A}-S^{th}_{A}].
\end{equation}
This large central charge limit result matches exactly with eq.(\ref{s39}) obtained through our covariant holographic entanglement negativity conjecture
from the bulk. Naturally this constitutes extremely strong evidence in favor of our conjecture for the $AdS_3/CFT_2$ scenario and indicates towards its universality.

\section{Covariant holographic entanglement negativity of a $CFT_{1+1}$ dual to a rotating extremal BTZ }
In this section we proceed to compute the holographic entanglement negativity of a mixed state for an extended bipartite quantum system described by a $CFT_{1+1}$ that is dual to extremal rotating BTZ black hole using our covariant prescription and demonstrate that for this case also the entanglement negativity captures the distillable quantum entanglement. The CFTs that are dual to the extremal black holes are very subtle and have many interesting properties. In \cite{Hartman:2008pb}, the authors propose the Kerr-CFT correspondence according to which $(d+1)$- dimensional extremal Kerr-AdS black holes are dual to a chiral half of a CFT in $d$ dimensions. In the $AdS_3/CFT_2$ context, it was shown in \cite{Caputa:2013lfa} that the entanglement entropy of a chiral half of a $CFT_{1+1}$ matches exactly with the covariant holographic entanglement entropy computed from the bulk BTZ black hole employing the HRT proposal. This was found to be true irrespective of whether one considers the near horizon metric or the full metric. The metric of the extremal BTZ black hole may be obtained by equating the inner and the outer radius of horizon i.e $r_{+}=r_{-}=r_0$ in eq.(\ref{s31}) as
\begin{equation}\label{s41}
 ds^{2}=-\frac{(r^2-r_{0}^2)^2}{r^2}dt^2 + \frac{r^2 dr^2}{(r^2-r_{0}^2)^2}+ r^2(d\phi-\frac{r_{0}^2}{r^2}dt)^2,
\end{equation}
where $J=M=\frac{r_0^2}{4G_N^{(3)}}$ from eq.(\ref{s32}).
 Note that the holographic entanglement entropy of a single interval in a $CFT_{1+1}$ dual to an extremal BTZ black hole was computed in \cite{Caputa:2013lfa} and we first briefly review their computation. In order to obtain the required geodesic lengths it is required to make the following coordinate transformation to map the above metric in eq.(\ref{s41}) to that of the pure $AdS_3$ space time in the Poincare coordinates 
\begin{eqnarray}\label{s42}
 w_{+}&=& \phi+t-\frac{r_0}{r^2-r_0^2},\label{s421}\\
 w_{-}&=&\frac{1}{2r_0}e^{2r_0(\phi-t)},\label{s422}\\
 Z&=&\frac{1}{\sqrt{r^2-r_0^2}}e^{r_0(\phi-t)}\label{s423}.
\end{eqnarray}

Note that these transformations can not be obtained by naively putting $r_{+}=r_{-}$ in eq.(\ref{s33}) and eq.(\ref{s33i}).
Under the transformations given by eq.(\ref{s421}), eq.(\ref{s422}) and eq.(\ref{s423}) the metric of the extremal BTZ black hole in eq.(\ref{s41}) becomes
\begin{equation}\label{s43}
 ds^2=\frac{dw_{+}dw_{-}+dZ^2}{Z^2}\equiv\frac{-dT^2+dX^2+dZ^2}{Z^2}.
\end{equation}

The computation of the length of the spacelike geodesic is similar to the case involving the non extremal black holes discussed earlier, leading to the following expression for the entanglement entropy 
\begin{eqnarray}\label{s44}
{\cal L}_{\gamma}&=&\log[\frac{(\Delta \phi)^2}{ \varepsilon_{i}\varepsilon_{j} }]\\
 \varepsilon_{i} &=& \frac{1}{r_{\infty}}e^{r_0(\phi_{i}-t_0)}.
\end{eqnarray}

Upon re-expressing the above equation in the BTZ coordinates and using the Ryu-Takayanagi formula ($S_{\gamma}=\frac{{\cal L}_{\gamma}}{4G_N^{(3)}}$), the entanglement entropy of a subsystem-$\gamma$ in the dual $CFT_{1+1}$ is given by the following expression (see \cite{Caputa:2013lfa} for details)
\begin{equation}\label{s45}
  S_{\gamma}=\frac{c}{6}\log\big[\frac{\mid\phi_{i}-\phi_{j}\mid}{a}\big]+\frac{c}{6}\log\big[\frac{1}{r_0a}\sinh\big(r_0\mid\phi_{i}-\phi_{j}\mid\big)\big].
\end{equation}

Note that the first term in the above equation resembles the entanglement entropy of a subsystem $\gamma$ of a zero temperature $CFT_{1+1}$ whereas the second term is the entanglement entropy of the same subsystem in a $CFT_{1+1}$ with an effective temperature $\frac{r_0}{\pi}$. The authors in \cite{Caputa:2013lfa} noted that this has a clear explanation that the left movers of the CFT are in ground state while the right movers have an effective temperature known as the Frolov-Thorne temperature\footnote{ The Frolov-Thorne temperature may be understood as follows. The left and the right moving temperatures of the dual $CFT_{1+1}$ may be expressed in terms of the inner and outer horizon radius of the bulk black string using eq .(\ref{s32}) i.e $\frac{1}{\beta_{+}}=T_{+}=\frac{r_{+}-r_{-}}{2\pi}$ and $\frac{1}{\beta_{-}}=T_{-}=\frac{r_{+}+r_{-}}{2\pi}$.  When black string becomes extremal the thermal temperature vanishes $T_+=0$ but there remains an effective temperature called the Frolov-Thorne temperature $T_{FT}=\frac{r_0}{\pi}=\frac{1}{\beta_{-}}$. This results in the entropy of the extremal black string in the bulk $s=\frac{ r_0}{4G_N^{(3)}} $ which in the dual $CFT_{1+1}$ corresponds to the thermodynamic entropy density  $s=\frac{\pi c}{6\beta_{-}}$. }\cite{PhysRevD.39.2125} given by 
\begin{equation}
 T_{FT}=\frac{1}{\beta_{-}}=\frac{r_0}{\pi}.
\end{equation}

Therefore, the entanglement entropy in eq.(\ref{s45}) may be re-expressed as
\begin{equation}\label{s45a}
  S_{\gamma}=\frac{c}{6}\log\big[\frac{\mid\phi_{i}-\phi_{j}\mid}{a}\big]+\frac{c}{6}\log\big[\frac{\beta_{FT} }{\pi a}\sinh\big(\frac{\pi\mid\phi_{i}-\phi_{j}\mid}{\beta_{FT}}\big)\big].
\end{equation}

Having obtained the required  expression for the lenght of a geodesic anchored to a given subsystem on the boundary, we identify the points $(\phi_1,\phi_2,\phi_3,\phi_4)\equiv(-L,u,v,L)$ and the subsystems $A\equiv[u,v]$, $B_1\equiv[-L,u]$ and $B_2\equiv[v,L]$ then we substitute all the quantities in the eq.(\ref{s37}) to obtain the covariant holographic entanglement negativity as
\begin{equation}\label{s46}
{\cal E}=\frac{c}{4}\log\big[\frac{\ell}{a}\big]+\frac{c}{4}\log\big[\frac{\beta_{FT} }{\pi a}\sinh\big(\frac{\pi\ell}{\beta_{FT}}\big)\big]-\frac{\pi c \ell}{4 \beta_{FT}}.
\end{equation}

For brevity the above expression may be re-written as
\begin{equation}\label{s47}
 {\cal E}=\frac{3}{2}\big[S_{A}-S^{FT}_{A}\big].
\end{equation}

Remarkably the above equation indicates towards an extremely interesting result that the covariant holographic entanglement negativity is the difference between the entanglement entropy $S_{A}$ and the thermodynamic entropy of the subsystem $A$ ( $S^{FT}_{A}=s\ell=\frac{\pi c \ell}{6\beta_{-}}$ ) suggesting that the latter does not contribute to the distillable quantum entanglement and behaves like an effective thermal entropy.
Note that this is because the ground state of the extremal black hole is highly degenerate giving rise to an emergent thermodynamic behavior. The extremal black hole entropy is the measure of this degeneracy. The dual $CFT_{1+1}$ is in a mixed quantum state having this counting entropy with an effective temperature equal to the Frolov-Thorne temperature \cite{Caputa:2013lfa}. Therefore the contributions from the counting entropy have to be eliminated to obtain the  distillable quantum entanglement which is measured by the entanglement negativity. This explains our result in eq.(\ref{s47}) through which we have demonstrated that once again the covariant holographic entanglement negativity of a bipartite system described by a $CFT_{1+1}$ dual to an extremal rotating BTZ black hole, computed employing our conjecture leads to the distillable quantum entanglement.

\section{Entanglement negativity in a $CFT_{1+1}$ with angular momentum and at zero temperature}
In this section we compute the entanglement negativity of a zero temperature $CFT_{1+1}$ with a conserved angular momentum and demonstrate that this matches exactly with the bulk  result obtained using our conjecture in the previous section. The authors in \cite{Caputa:2013lfa} determine the conformal transformation that maps the points on the complex plane to that on the cylinder where the dual CFT is situated by observing how the co-ordinate transformations in eq.(\ref{s421}) and eq.(\ref{s422}) behave as $r\rightarrow \infty$. The authors then use the transformation to show that the entanglement entropy computed from the two point function of the twist and anti twist fields on such a cylinder matches exactly with the holographic entanglement entropy obtained from the bulk extremal BTZ black hole. The behavior of eq.(\ref{s421}) and eq.(\ref{s422}) as $r\rightarrow \infty$ leads to the following conformal map 
\begin{eqnarray}
 w&=&z,\label{s51}\\
 \bar{w} &=& \frac{\beta_{-}}{2\pi}\log[\frac{2\pi \bar{z}}{\beta_{-}}].\label{s52}
\end{eqnarray}

As discussed in section-4, from eq.(\ref{s26}) and eq.(\ref{s25}), the four point function in the $CFT_{1+1}$ on the cylinder is related that on the complex plane by
\begin{eqnarray}
 \big<{\cal T}_{n_e}(w_1)\overline{{\cal T}}^2_{n_e}(w_2){\cal T}^2_{n_e}(w_3)\overline{{\cal T}}_{n_e}(w_4)\big>_{\beta_{-}}&=&\prod_{i=1}^{4}\mid\frac{dz_i}{dw_i}\mid^{\Delta^{i}_{n_{e}}} \big<{\cal T}_{n_e}(z_1)\overline{{\cal T}}^2_{n_e}(z_2){\cal T}^2_{n_e}(z_3)\overline{{\cal T}}_{n_e}(z_4)\big>_{\mathbb{C}},\nonumber\\ 
 &=&\prod_{i=1}^{4}(\frac{dz_i}{dw_i}\frac{d\bar{z_i}}{d\bar{w_i}})^{\frac{\Delta^{i}_{n_{e}}}{2}}\frac{c_{n_e}c^{(2)}_{n_e}}{z_{14}^{2\Delta_{n_e}}z_{23}^{2\Delta^{(2)}_{n_e}}}\frac{{\cal F}_{n_e}(x)}{x^{\Delta^{(2)}_{n_e}}}.\label{s53i}
\end{eqnarray}

We identify the points at which the four point function has to be evaluated  as $(w_1,w_2,w_3,w_4)\equiv(-L,u,v,L)$, therefore from eq.(\ref{s52})  we have $(z_1,z_2,z_3,z_4)\equiv(-L,u,v,L)$ and $(\bar{z}_{1},\bar{z}_{2},\bar{z}_{3},\bar{z}_{4})\equiv\frac{\beta_{-}}{2\pi}(e^{-\frac{2\pi L}{\beta_{-}}},e^{\frac{2\pi u}{\beta_{-}}},e^{\frac{2\pi v}{\beta_{-}}},e^{\frac{2\pi L}{\beta_{-}}})$. After this identification, we compute the four point function in eq.(\ref{s53i}) and substitute it in eq.(\ref{r19}) to obtain the entanglement negativity as
\begin{equation}\label{s54}
\begin{split}
 {\cal E}=\frac{c}{4}\log\big[\frac{\ell}{a}]+\frac{c}{4}\log\big[\frac{\beta_{-}}{\pi a}\sinh{(\frac{\pi\ell}{\beta_-}})\big]-\frac{\pi c\ell}{4\beta_-}+f(e^{-\frac{\pi \ell}{\beta_-}})+\ln[c^2_{1/2} c_1].
 \end{split}
\end{equation}

As discussed in section-5, in the large-$c$ limit the entanglement negativity receives the leading contribution from the universal terms which of $O[c]$ (first three terms in the above equation) and the non universal terms which include the function $f(x)$ and the constant are subleading. Furthermore, identifying  $\beta_-=\beta_{FT}$ we obtain the large-$c$ limit of the entanglement negativity of a zero temperature $CFT_{1+1}$ with a conserved angular momentum as
\begin{equation}\label{s55}
 {\cal E}=\frac{3}{2}\big[S_{A}-S^{FT}_{A}\big].
\end{equation}
The above expression for the entanglement negativity exactly matches with eq.(\ref{s46}) obtained from the dual bulk extremal rotating BTZ black hole using our covariant holographic entanglement negativity conjecture. Thus,  we have verified our conjecture by demonstrating that in the large central charge limit it exactly reproduces the entanglement negativity of a mixed state for an extended bipartite system described by a $CFT_{1+1}$ dual to a stationary $AdS_3$ configuration involving a bulk extremal rotating BTZ black hole. 
 
\section{Summary and Conclusion}

To summarize we have proposed a covariant holographic conjecture for the entanglement negativity of mixed states in bipartite quantum systems characterized by $d$ dimensional conformal field theories dual to bulk non-static $AdS_{d+1}$ configurations. Employing our conjecture we have evaluated 
the entanglement negativity of a mixed state for an extended bipartite quantum system described by a finite temperature $(1+1)$ dimensional holographic CFT dual to a bulk rotating non extremal BTZ black hole. Remarkably in the large central charge limit this exactly matches our computation for the entanglement negativity of the above $1+1$ dimensional holographic CFT through the replica technique. Furthermore as expected from quantum information theory considerations we observe that the entanglement negativity involves the elimination of the thermal contributions and hence leads to an upper bound for the distillable entanglement. Interestingly on the other hand our conjecture applied to the case of zero temperature $1+1$ dimensional holographic CFTs dual to extremal rotating BTZ black holes lead to the entanglement negativity of the chiral half and involves the elimination of the corresponding Frolov-Thorne entropy.

The covariant holographic entanglement negativity conjecture proposed by us in this article provides an elegant holographic scheme for exploring entanglement issues for mixed states involving time dependent processes in holographic conformal field theories. Naturally this should find interesting applications 
in diverse areas such as quantum quenches and thermalization, quantum phase transitions which involve entanglement evolution in condensed matter systems.
Furthermore our covariant holographic entanglement negativity conjecture is expected to find application to the study of the long standing issue of black hole formation and the information loss paradox and the related firewall problem. These would be extremely interesting open issues for future investigations.

\section {Acknowledgement }
The work of Pankaj Chaturvedi is supported by  Grant No. 09/092(0846)/2012-EMR-I, from the Council of Scientific and Industrial Research (CSIR), India.

\bibliographystyle{utphys} 
\bibliography{CoHEN}

\end{document}